\documentclass[prb,preprint]{revtex4-1} % style for Physical Review B and AJP are similar

\usepackage{amsmath} % need for subequations

\usepackage{graphicx} % for figures

\begin{document}

\title{Energy, Entropy, Information, and Intelligence}
\author{Vladim\'\i r \v Cern\'y}
\affiliation{Department of Theoretical Physics and Physics Education, Faculty of Mathematics, Physics and Informatics, Comenius University, Bratislava, Slovakia}
\email{cerny@fmph.uniba.sk} %optional

\date{\today}

\begin{abstract}
The paper presents a lightweight discussion of relations between energy, entropy, information, and intelligence, based on an analysis of the energy needed for computation.
\end{abstract}

\maketitle

\section{Introduction}
This paper is a remake and an extrapolation of ideas presented in the chapter ``Reversible computation and the thermodynamics of computing'' in a beautiful Feynman book.\cite{Feynman} Basically,  it is about energetics of computing. A computer is a physical device, whose dynamics we use for computing. Therefore the question on energy requirements is certainly relevant. This question has been with us since the time of von Neumann, when computers were just born. The discussion on minimal energy required for computing culminated in 1960's and 70's in the works of Landauer\cite{Landauer}, Bennett\cite{Bennett}, Toffoli\cite{Toffoli} and others. The review by Bennett\cite{Bennettrev} can be consulted for details. Finally, the discussion converged on the recognition that reversible dynamics can be used for computing purposes and on the fact that when used, energy needs to be dissipated only during the process of memory resetting (sometimes the term ``erasure of information'' is used, even though it may be confusing). We shall repeat the Feynman argument why energy is needed to reset memory into the default state in Section~\ref{sec:erasure}. We shall also dwell on discussion of a complementary process: how information can be used to extract useful work. The line of reasoning will proceed along the following points.
\begin{itemize}
  \item Resetting memory requires energy.
  \item Energy needed for memory resetting is proportional to the amount of erased information.
  \item A device able to perform alternative actions needs memory resetting to be reusable.
  \item Information (knowledge) can be used to extract useful work.
  \item A device able to extract ``work from information'' needs to spend the same amount of work to be reused.
  \item Intelligent inductive inference (plus a good deal of luck) enables to extract useful work without the need for resetting.
\end{itemize}
\section{Resetting memory requires energy}\label{sec:erasure}
\begin{figure}[h!]
\centering
\includegraphics[width=4.0 cm]{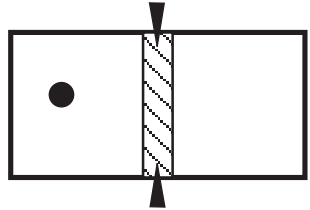}
\caption{Feynman Register.}
\label{fig:Freg}
\end{figure}

\begin{figure}[h!]
\centering
\includegraphics[width=4.0 cm]{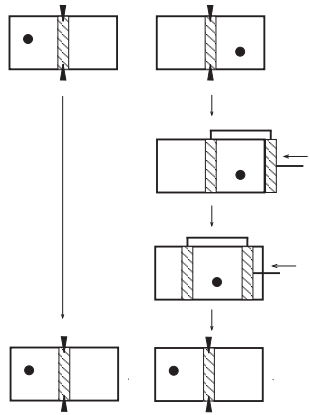}
\caption{Resetting to the default state from a known state (qualified resetting).}
\label{fig:reset1}
\end{figure}

Feynman invented a simple physical model of a one-bit register, presented in Fig.~\ref{fig:Freg}. It is a cylinder divided into two compartments by a central partition. The cylinder contains just one molecule, which can be either in the left compartment (representing logical 0) or in the right compartment (representing logical 1). Let us define the default register state to be ``0'' (the molecule in the left compartment).
It is not without interest that when we know the actual state of the register, we can bring it to the default state ``0''  (reset it) without spending energy. The process of resetting a register from a known state (we shall call it qualified resetting) is graphically presented in Fig.~\ref{fig:reset1}. If the known actual state is ``0'', we need not do anything. If the known state is ``1'', we use the central partition as a piston, joining it firmly with the piston attached to the cylinder from the right side and move the two pistons together to the left end. This operation does not require performing mechanical work from our side, since the forces of the (one-molecule) gas acting on the two pistons are opposite and equal in size. Finally we convert the piston in the central position into a partition and close the cylinder  at the right end. Voila, the register is in the default state and it has cost no energy. This is not the end of the story, we shall add an important discussion in Section~\ref{sec:qualified}.

\begin{figure}[h!]
\centering
\includegraphics[width=4.0 cm]{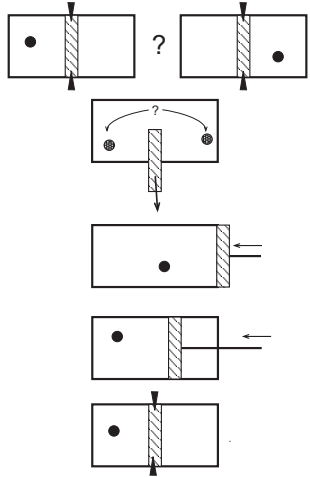}
\caption{Resetting to the default state from an unknown state (unqualified resetting).}
\label{fig:reset}
\end{figure}

Now let us discuss how to bring the register to the default state if we do not know its actual state (unqualified resetting). The method is presented in Fig.~\ref{fig:reset}.
We start by pulling the central partition aside. Then pushing a piston from the right side until the central position brings the molecule into the left compartment. For this, however, we need to perform mechanical work, since the (one-molecule) gas acts on the piston by its pressure $p$. The calculation of the needed amount of work $W$ is easy.
\begin{equation}
\label{eq:compression}
W = -\int_{V}^{V/2}pdV = -\int_{V}^{V/2}\frac{kT}{V}=-kT (\ln(V/2)-\ln(V))=kT\ln 2.
\end{equation}
Some notes: If we were calculating the work performed by the gas, there would be no minus sign. We are, however, calculating the work performed by an external hobbit pushing the piston, acting by force opposite in direction to that of the gas force and therefore the minus sign. We have used the gas state equation for $N=1$ molecules: $pV=NkT$.

What happens with the work $kT\ln 2$? We have effectively assumed that the cylinder is in contact with the environment having temperature $T$. In order not to increase the gas temperature, the energy $kT\ln 2$ has to be dissipated into the environment as heat.

Summarizing: we have shown that erasing (destroying) 1 bit of information costs energy $kT\ln 2$ to be dissipated into environment. We did not discuss in detail any other operation with memory registers like setting the register into desired memory state starting from the default state  or copying the state of some register to another register, but it is known that those operations do not require performing work.\cite{Feynman}

\section{Information, entropy, energy}

In the previous section, we have manipulated gas of a single molecule. The results extrapolated to gas of $N$ molecules bring a new insight into the problem. We have seen that to compress gas, we have to perform work. On the other hand, if gas is expanding isothermally, it can perform useful mechanical work. Similar calculation as in Eq.~\ref{eq:compression} gives the amount of work \emph{performed by gas} doubling its volume as
\begin{equation}\label{eq:work}
W'=NkT\ln 2.
\end{equation}

Somebody has to pay for this work: the environment in the form of heat
\begin{equation}
\label{eq:heat}
Q'=NkT\ln 2 .
\end{equation}

However, during the gas expansion something else happens behind the scenes: our knowledge about the gas microstate decreases. Let us discuss it in more detail. Gas state is macroscopically given by a few parameters, such as  pressure $p$, volume $V$ and temperature $T$. We say we have specified the gas macrostate. Microscopically the gas state is a tremendously complex notion. Describing the microscopic state requires giving the position and velocity of each individual molecule (we neglect quantum-mechanical complications). So knowing the macrostate, there is a good deal of information we are missing about the actual gas microstate. Now we shall argue that by expansion the amount of this missing information increases.

\begin{figure}[h!]
\centering
\includegraphics[width=5.0 cm]{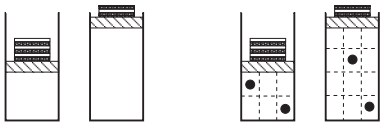}
\caption{The increase of the number of microstates by expansion.}
\label{fig:gasexpansion}
\end{figure}

Let us consider for simplicity a two-molecule gas that is doubling (isothermally and reversibly) its volume as presented in Fig.~\ref{fig:gasexpansion}.
At the left side of the figure the original and final states are presented as macrostates. At the right side those states are presented as microstates. For simplicity we neglect molecular velocities, so a microstate is given by positions of all the molecules. In order to record positions of molecules, the gas volume is divided into smaller (virtual) compartments, six are presented (as an oversimplified example) for the original state, 12 for the final state. In the original state placing two molecules into 6 compartments is possible by $(6\times5)/2=15$ ways. In the final state the number of possible microstates is $(12\times 11)/2=66$. So to communicate the exact microstate for the macrostate with smaller volume one has to specify a number between 1 and 15, to communicate the microstate for the doubled volume one has to specify a number between 1 and 66. Since there are more possibilities for the large volume, it is natural to say that if we do not know the exact microstate, the missing information is larger for larger volume.

It is not difficult to make the statement more quantitative. We can play the same game with virtual compartments, just their number $C$ must be very very large. For $N$ molecules in $C$  compartments the number $n$ of possible microstates is
\begin{equation}
\label{eq:n}
n=\frac{C!}{(C-N)!N!} .
\end{equation}
If I want to communicate to you which microstate actually realizes the macrostate you observe, I have to send you its id number $i$: a number between $1$ and $n$. The message would have $\approx\log_2 i$ bits, since that is the number of binary digits representing the number $i$. The mean number of bits sent will be
\begin{equation}
\label{eq:I}
I=\overline{\log_2 i}=\frac{1}{n}\sum_{i=1}^n \log_2 i=\frac{1}{n}\log_2 n!=\frac{1}{n}(\log_2\mbox{e})\ln n!
\end{equation}
For
\begin{equation}
C\gg N\gg 1
\end{equation}
we can use (several times) Stirling approximation $\ln n!\approx n \ln n - n$ and combining Eqs.~\eqref{eq:n} and \eqref{eq:I} we get
\begin{equation}
I\approx N (\log_2\mbox{e}) \ln \frac{C}{N}
\end{equation}
If we decide to measure the information content of a message by its number of bits (we can do that if the coding used was optimal), we can conclude that if we know just the gas macrostate and not the microstate, we are missing information of I bits on average. Physicists prefer to measure information not in bits but in nats just omitting the factor $\log_2\mbox{e}$. Somewhat surprisingly, however, physicists found, that they already had defined a phenomenological physical quantity, the entropy $S$, which differs from $I$ expressed in nats only by another information-unit-changing factor $k$ (the Boltzmann constant). So finally we get for the gas entropy
\begin{equation}
S\approx k N \ln \frac{C}{N} + \mbox{term which takes into account molecular velocities}
\end{equation}
To repeat: $S$ measures the amount of missing information (in units nats~$\times$~Joule/Kelvin) if we know only the gas macrostate and not its microstate.

So let us come back to the problem of expanding gas. Doubling the volume means doubling the number of compartments and therefore an increase of entropy
\begin{equation}\label{eq:deltaS}
    \Delta S =k N \ln \frac{2C}{N} - k N \ln \frac{C}{N} =k N \ln 2
\end{equation}
The terms corresponding to molecular velocities canceled out, since the process is isothermal, and statistically the molecular velocities do not change. Increasing the volume lead to two effects
\begin{itemize}
  \item extracting heat from the environment and changing it into useful work, and
  \item increasing the amount of missing information or, equivalently, decreasing the amount of our information about the gas microstate
\end{itemize}

So we see that decreasing the amount of our knowledge enables us to extract heat \eqref{eq:heat} from the environment and change it into useful work \eqref{eq:work}. We shall discuss it in details in later sections.

Comparing equations \eqref{eq:deltaS} and \eqref{eq:heat} we find the relation
\begin{equation}\label{eq:QS}
    Q= T \Delta S
\end{equation}
Equation \eqref{eq:QS} is generally valid for any reversible process in nature, we have observed just one very special case.

Now we understand why we have to pay by performing work and dissipating heat in the process of unqualified resetting of a memory register into the default state (see Equation \eqref{eq:compression}). Before the proces of memory reset we did not know the gas state (the position of the molecule), after the proces we know that the register is in the default state (logical 0, molecule in the left compartment). So the amount of our knowledge increased. Therefore the entropy decreased, $\Delta S < 0$ so $Q = T \Delta S < 0$. So the acquired heat is negative: heat has been dissipated.

\section{Ability to perform (repeatedly) alternative tasks requires energy dissipation}
\label{sec:qualified}

We have shown that qualified resetting of memory requires no work, while unqualified resetting of one bit of memory requires dissipation of heat $Q=kT \ln 2$. A question might arise: why do we usually perform unqualified memory resetting ``as if'' from an unknown state? In many cases, the register to be reset contains the result of the calculation. Most probably we have printed its content on a piece of paper, so we know its state and we should be able to do qualified resetting.

The reason is that the device able to perform only unqualified memory resetting is simpler than the device able to perform qualified memory resetting. Look again at the Figures \ref{fig:reset1}~and~\ref{fig:reset}. The unqualified resetting can be performed by a ``single-purpose hobbit''. The hobbit is trained to come from the right side to the register and push the piston. The qualified resetting needs a hobbit able to perform two alternative tasks. One is ``do not do anything'', the other is ``bind the pistons and push them from the right''.

\begin{figure}[h!]
\centering
\includegraphics[width=4.0 cm]{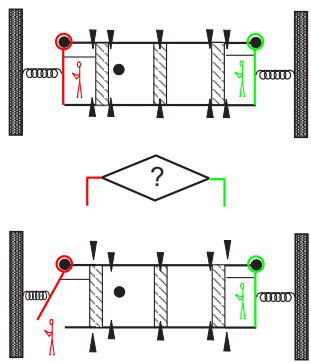}
\caption{A general mechanical device enabling to harness two unqualified hobbits to perform qualified work. In the upper part the device is in the idle state, all the pistons are blocked, the hobbits are confined in their cages. In the lower part the device was put to action, the pistons were unblocked. Now one of the two cages is opened depending on the state of the internal Feynman memory register.}
\label{fig:hobbits}
\end{figure}

We shall argue that, quite generally, the ``qualified hobbit'' needs to have internal memory register telling him which of the two alternatives to perform. In Fig.~\ref{fig:hobbits} we present a mechanical device (programable ``processor'')
harnessing two ``unqualified hobbits'' to perform alternative works. There are two cages (left and right) with two unqualified hobbits in them. The doors of the cages can be manipulated by the pistons, which feel the pressure of a Feynman one-molecule memory register.

Let us describe how the device is operated by a supervisor. If the device is in the idle state, than all the pistons are blocked, and the hobbits are confined in their cages. Now the supervisor decides to use the device to perform work for which the left hobbit is trained. He has to set the internal Feynman memory register of the device to the ``0'' state (this is how the supervisor programs the device) and to unblock all the pistons.
If the register is in the state ``0'', then only the left piston feels the pressure. It moves just a little, opens slightly the door of the left cage. The left hobbit escapes through the small opening and performs its task (and is trained to come back into the cage as a part of his ``single-purpose algorithm''). The ``working place'' of the hobbit is not depicted in the Fig.~\ref{fig:hobbits}. For example if the device is used to perform a qualified reset of some register, the hobbit leaves his cage, walks to the register (not depicted in the figure), performs the work he is trained for and returns back to his cage. Were the internal register in state ``1'', it would be the right hobbit to perform his task. We have shown that a one-bit internal memory register can be used (by the supervisor) to set which of the two alternatives to perform. Since the doors are opened only slightly, only infinitesimal work is performed by the gas and this can be neglected. Therefore we do not need additional work to perform alternative tasks.

But wait a moment! After the work was done, both hobbits are back in their cages, the pistons are blocked, the device is in the idle state. However, the internal register remains ``armed for the left''. If the supervisor does nothing with that, the machine effectively becomes a ``single purpose'' (unqualified) device performing in each work cycle just the ``left algorithm''. If the supervisor wants to use the device as a qualified device in the next work cycle, he has to reset its internal register. After resetting, the register is prepared to be programmed by copying new information into it. Unqualified resetting costs work, copying new information needs no additional energy costs. We stress that the register reset should be done in an unqualified way, otherwise additional qualified-work-able device should be used for the resetting. This additional device has to have its internal register which itself needs to be reset after work and the chain would grow indefinitely unless at some stage an unqualified reset is used.

Here is the point where a thorough reader might stop and suggest: why the hobbits do not reset their supervising register themselves (as the last part of their work). The left hobbit will do nothing (the register being in the default state), the right hobbit will do the two-joined-pistons-push operation (see Fig.\ref{fig:reset1}). Qualified resetting done, no energy dissipation needed! Unfortunately, this cannot be done. The discussion is presented in the Appendix.

Now we take courage and generalize the result. We conjecture that any device able to perform repeatedly  qualified (two alternatives) work has to contain an internal one-bit memory register, which should be reset in an unqualified way. This needs heat $Q=kT\ln 2$ to be dissipated into the environment for each working cycle. We cannot use qualified resetting: for that we would need another qualified device which itself should have internal register to be reset, so we would just push the energy expenditure one step further.

Conclusion: memory reset needs to be done in an unqualified way and costs energy $kT\ln 2$ per bit. It is worth to stress that once the internal register of the device is in the default state, no work is needed to set it into desired state or copy to it the information content of some other register. So ``programming the qualified hobbit'' costs zero energy.

\section{Is it possible to ``extract work from information''?}

In this section we shall discuss the opposite process to memory resetting:``extracting work from information''. The original idea is due to Bennet\cite{Bennettrev} (``a tape full of zeros can act as a `fuel' doing useful thermodynamic or mechanical work as it randomizes itself'').

\begin{figure}[h!]
\centering
\includegraphics[width=4.0 cm]{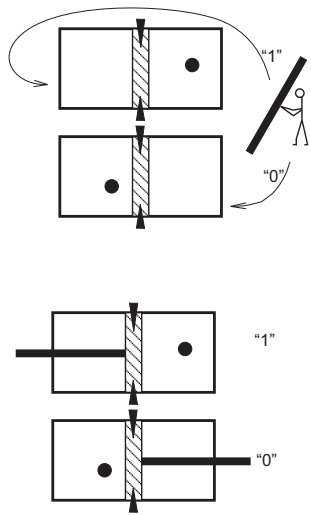}
\caption{Preparations to extract work.}
\label{fig:ExA}
\end{figure}

\begin{figure}[h!]
\centering
\includegraphics[width=4.0 cm]{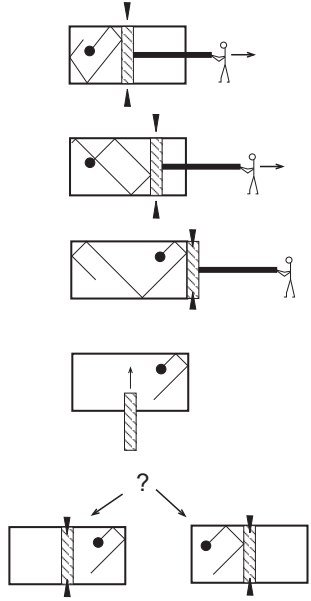}
\caption{Process of work extraction leading to loss of information.}
\label{fig:ExB}
\end{figure}

The principle is explained in Figs.~\ref{fig:ExA} and \ref{fig:ExB}.
Suppose I have a one-bit register and I know its state. I call a hobbit and send him to the register from the right side if the register is in state ``0'' and from the left side if it is in state ``1''. The hobbit mounts his handle to the central partition making a piston of it. Then I unblock the partition and the gas starts to expand performing work. The hobbit is consuming that work using it for any useful purpose. When the partition-piston arrives at the end of the cylinder and I insert a new partition to the central position (inserting partition does not cost any work!), I do not know at all in which compartment (left or right) the molecule is caught. The initial information was used to extract useful work, while the register was randomized. So the entropy of the register increased by
\begin{equation}
    \Delta S = k\ln 2
\end{equation}
the hobbit extracted the work
\begin{equation}
    W =kT\ln 2
\end{equation}
while the environment payed for that by heat
\begin{equation}
    Q =kT\ln 2
\end{equation}
used to keep the temperature constant. So the information helped me to exchange a chaotic form of energy (through heat) for useful (``organized'') mechanical work.
\begin{figure}[h!]
\centering
\includegraphics[width=7.0 cm]{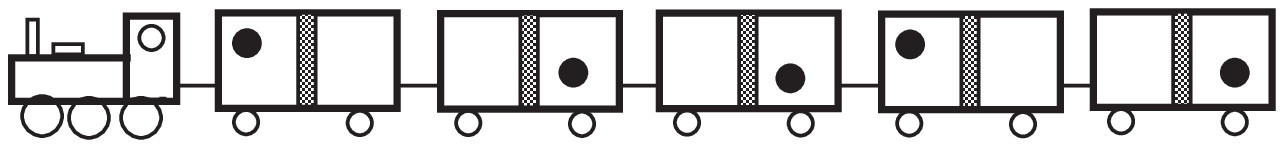}
\caption{Sending information as a demonstration of the famous Landauer's quotation ``information is physical''.}
\label{fig:Train}
\end{figure}
So if somebody sends me the whole train of N registers (see Fig.~\ref{fig:Train}) together with a delivery note stating for each register its memory state, I can extract from it the work
\begin{equation}
    W =N kT\ln 2.
\end{equation}
Good news!

But wait a moment! To extract work from the train I need an army of qualified hobbits! Qualified, since they must be able to perform alternative procedures (come to the register either from the left side or the right side). We have learned already that for that they need to have memory registers ``in their heads'', instructing them either to go to the left or the right. If I want to use the same hobbits to unload the next train, I have to reset their registers, and for that I have to pay by spending the work
\begin{equation}
    W =NkT\ln 2
\end{equation}
This is exactly the same work which I got by unloading the previous train. The net result is just zero. Bad news!

\begin{figure}[h!]
\centering
\includegraphics[width=7.0 cm]{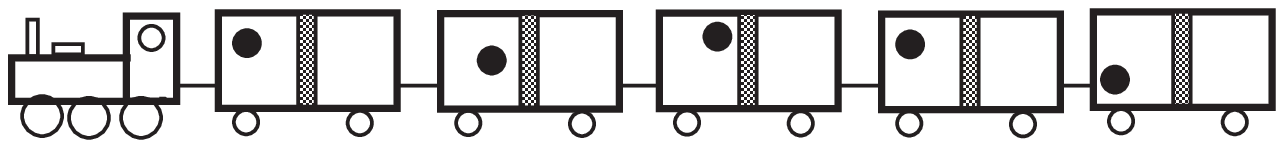}
\caption{Uniform registers providing extractable energy source.}
\label{fig:Train0}
\end{figure}

There seems to be a way out. My energy supplier has to send trains (Fig.~\ref{fig:Train0}) containing N registers each in the same memory state, for example ``0''. (This is what Bennett was saying in the quotation at the beginning of this section.) To extract energy, unqualified (single-purpose) hobbits are sufficient. Such hobbits do not have a register to be reset after work, so net work gain is positive. Good news!

But wait a moment! Now it is the energy supplier who might have energetic problems. He has to load the train ``with zeros exclusively''. If he has access only to a pile of registers in random states, he has to check (measure) the state of each register and then to load it or not to load it according to the findings. This means alternative processes: to load or not to load (yes, Hamlet also needs a register!), so now the energy supplier needs qualified hobbits with registers needing reset. So the supplier spends the same amount of work as it is gained at the side of a customer. So the total net amount is again zero. Bad news!

But wait a moment! There is a chance described in the next section.

\section{Intelligence and inductive inference}

The reader should be warned that this section contains really a lightweight discussion. It is more entertainment then science, although I believe deeper aspects are hidden behind the scenes.

We start our discussion with a seemingly unrelated topics: tests of intelligence. (Well, that is a very controversial subject itself.) The first (and easiest) question one meets in the intelligence tests is the following

2, 4, 6, 8, ... What is the next number in the sequence?

If you answer 2, 4, 6, 8, 10 you pass the test. If you instead write
\begin{equation}
    2, 4, 6, 8, 34
\end{equation}
you fail. But why, actually? Maybe you have realized, that the formula
\begin{equation}\label{complicated}
    n^4-10n^3+35n^2-48n+24
\end{equation}
reproduces the first four numbers and the extrapolation gives 34! So maybe you are more intelligent than some other candidate with his rather simple-minded formula
\begin{equation}\label{simple}
    2n.
\end{equation}
(Clever people failing in intelligence tests stress this as a controversial matter for sure.)
The test tries to evaluate your ability of inductive inference, and prefers simple solutions to more complicated ones. The simple solutions are considered to be more intelligent. Maybe there is a good reason for that but instead of trying to understand formal theory of inductive inference\cite{Solomonoff}, let us for simplicity conjecture that this preference is
of Darwinian origin. To survive in a jungle, you must be able to find quickly solutions to problems you meet. If your mother teaches you a simple rule ``big mane, run away'' it helps you to survive better than if she gives you a copy of the complete lion genome to identify the animal. Relying on simple solutions, you need luck to survive. With complicated solutions you would not survive for sure. So maybe we are offsprings of beings relying on simple inductive inference and that is why we prefer simple solutions to more complicated ones.

Similarly, in a business jungle, if you make some ten control drills and each time you find a coal layer, you would consider to start a cole mine business. You simply bet that the Creator is not joking by distributing there just ten buckets of cole. Well, a bet is an hazard, the inductive inference is also a sort of an hazard, but what else you can do?

So let us come back to our energy supplier that wants to make money by loading trains with Feynman memory registers in state ``0''. What he can do is to prospect for piles of Feynman registers and take a few samples from each pile found. If all the samples from a particular pile would be ``state 0'', he can infer ``this is a pile of uniform Feynman registers in state 0''. He hires an army of cheap unqualified hobbits (no need to reset their registers!) and starts to deliver uniform (hopefully) information trains.  With a piece of luck the net total result would be positive extracted useful work.

\section{Conclusion}
Bet on intelligent inductive inference. Hopefully you will survive.

\appendix*

\section{Impossibility of self-resetting}

Here we discuss the idea suggested in Section \ref{sec:qualified} that the two hobbits (``left`` and ``right'') could perhaps reset their supervising register themselves. Well, here we pay the prize for the lightweightness of our discussion. We used hobbits to make the explanations comprehensible. By that, however, we abandoned scientific rigor. Behind the scenes everywhere in our discussion is a hidden assumption: hopefully one can substitute the lightweight argument by a rigor one.

Hobbits escaping from cages performing some work should somehow correspond to strictly reversible-dynamics realization of the computing tasks. Reversibility is essential if we do not want to dissipate additional energy by the ``working hobbits''. We did not even try to support this hope for reversible realization. If reversible implementation were not possible, it would just mean that we have to dissipate even more energy, what does not spoil our argument that the ability to perform (repeatedly) alternative tasks requires energy dissipation.

\begin{figure}[h!]
\centering
\includegraphics[width=7.0 cm]{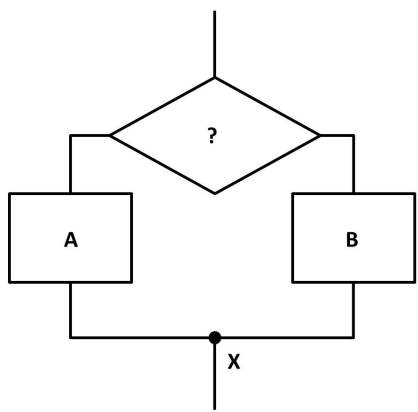}
\caption{Flowchart of the self-resetting scheme}
\label{fig:self}
\end{figure}

We shall argue now that the self-resetting-hobbit scheme, though being so attractive, is not reversibly realizable. The essence of our argument can be found in Bennett's paper.\cite{BennetReply} The flow chart of the self resetting scheme is presented in Fig.~\ref{fig:self}. In the point X the hobbits are back in their cages and the register is reset. That means that in the point C we cannot tell, whether we have got there through the branch A or B. It means that the scheme is logically irreversible (we cannot ``undo'' the calculation). Logical irreversibility means dynamical irreversibility, what means energy dissipation. Self resetting is not possible without energy dissipation.

\begin{acknowledgments}
Encouraging discussions with Vlado Balek and Pavol Cerny which helped to improve our argumentation are highly acknowledged. In particular they pointed at the problem of ``self-resetting'' as discussed in the Appendix. Special thanks to Pavol Brunovsk\'y for suggesting that hobbit is a nicer word than agent or dwarf.
\end{acknowledgments}

\end{document}